\documentclass[10pt]{article}
\usepackage[OE]{express}

\begin{document}

\title{ Mechanism of laser induced filamentation  in dielectrics}

\author{N. Naseri$^{1,2}$, G. Dupras$^{1}$, and L. Ramunno$^{1}$}

\address{$^1$ Department of Physics and Centre For Research In Photonics, University of Ottawa,
   ,25 Templeton St, Room 350
Ottawa, ON, K1N 6N5, Canada}
   
   \address{$^2$ Physics and Astronomy Department, Middle Tennessee State University, Wiser-Patten Science Hall, 422 Old Main Cir, Murfreesboro, TN 37132, U.S.A}




\begin{abstract}
Femtosecond laser filamentation in transparent media has a wide range of applications, from three dimensional manufacturing to biological technologies to supercontinuum generation. While there has been extensive investigations over the last two decades, there  remain aspects that are not understood, owing to the complexity of the interaction. We revisit intense femtosecond laser interaction with dielectric materials at 800nm via high resolution three dimensional simulations, where the complete set of Maxwell's equations is solved. We simulate filament formation for a range focusing conditions (including tight focusing) and for a range of laser energies, and through this are able to shed new insight on the dynamics. This includes the formation of two distinct damage zones for intermediate focusing, similar to what was seen but not fully understood almost two decades ago.
\end{abstract}



\section{Introduction}
Ultrafast laser filamentation in gasses, liquids, and solids has been an active research area since the advent of femtosecond lasers \cite{Braun,Couairon2007,Berge,Chin}. Filamentation in air has been widely investigated\cite{Chin2012}, with many applications ranging from atmospheric sensing \cite{Kasparian,Chin2009} to terahertz generation\cite{Kim}. While the critical power of self-focusing is in the gigawatt range for gasses, it is in the megawatt range for condensed matter due to their much higher nonlinear index, larger by up to three orders of magnitude \cite{Couairon2007}. Therefore, filamentation in liquids\cite{Liu,Potemkin} and solids\cite{Dubietis}  is interesting not only for many applications, but also because the physical mechanisms of filamentation have proven to be largely universal, and thus can be studied on a reduced scale.

Intense femtosecond laser interaction with transparent media was investigated over two decades ago in the context of material damage thresholds\cite{Stuart1995,Stuart1996} and for applications including fabrication of waveguides within bulk silica\cite{Davis}, gratings within optical fibers \cite{Kando1999}, three dimensional data storage \cite{Glezer1996,Watanabe}, tissue ablation \cite{Loesel} and supercontinuum generation \cite{Zozulya}; femtosecond filamentation within bulk dielectrics was also observed \cite{Tzor,Sudrie2002}.  Femtosecond laser processing via filamentation remains an active field, with ever greater precision being achieved through spatio-temporal tailoring of the input beam, including the use of Bessel and Airy beams, chirped pulses and bursts \cite{Bhuyan2010,Bhuyan2014,Xie2015,Mish,Erden,Courv}. Further, a new regime of filamentation in transparent media has been observed in the mid-IR, within the anomalous group velocity dispersion regime\cite{Smetanina,Durand}, leading to, for example, unprecedented multi-octave supercontinuum generation\cite{Silva}.

Despite this extensive body of work, there remain aspects of intense laser dielectric interaction and filamentation that are not understood. The majority of simulations are based on an equation for nonlinear pulse propagation with plasma generation and dynamics, where approximations such as slowly varying envelopes, or slowly evolving waves, among others, are applied to obtain an evolution equation for the pulse\cite{Couairon2011}. This approach, in general, has been highly successful when the correct approximations are considered for the problem at hand\cite{Dubietis}. However, filament properties can sometimes be difficult to predict due to rich and complex dynamics leading to the formation of complex structures such as, for example, self-assembled nanogratings \cite{Bhardwaj}. Thus, a more rigorous computational approach is sometimes required \cite{Popov,Bulgakova,Rudenko}.

In this paper, we revisit filamentation in silica at 800 nm with high resolution three-dimensional finite-difference time domain (FDTD) simulations, where the complete set of Maxwell's equations is solved. We report on filament formation for a range focusing conditions (including tight focusing from a parabolic mirror \cite{Popov2009}) and for various laser energies, in order to detangle which mechanisms are at play in each regime. Experimental studies of single-pulse filamentation in transparent material in the past two decades have shown that filamentation strongly depends on laser focusing conditions, from voids for tight focusing \cite{Glezer1997,Schaffer2001} to long channels of modified refractive index for loose focusing\cite{Yamada2001,Wu2003}. Through our detailed simulations, we are able to shed new insight on the dynamics of the intense laser-matter interaction, including the formation of two distinct damage zones for intermediate focusing, similar to what was seen but not fully understood almost two decades ago \cite{Sudrie2002,Coua2005}.

This paper is organized as follows: First we present our numerical model along with parameters used in our simulations. Then we investigate the physical mechanism leading to filamentation in fused silica for different focusing conditions at fixed laser energy. This is followed by the presentation of an analytical model for filamentation in solids based on that proposed by Lim et. al. \cite{Lim2014} for filamentation in air, to visualize the transition from Kerr self-focusing to geometrical focusing as laser focusing becomes tighter. Finally, we study the effect on filamentation of the laser intensity (power) for fixed laser spot sizes.

\section{Numerical Method} \label{NM}

Maxwell's equations are solved using the FDTD method\cite{Taflove2005} via the Yee algorithm\cite{Yee,Popov}  with the constiutive  relations (cgs units) $
  \textbf{H}=\textbf{B},
~
  \textbf{D}=(1+4\pi(\xi_l+\xi_kE^2)\textbf{E}$
and  current density
  $\textbf{J}=\textbf{J}_p+\textbf{J}_{PA}$.  $\textbf{E}$ and $\textbf{B}$ are the electromagnetic fields, $\textbf{D}$ the displacement vector, $\textbf{H}$  the magnetic field auxiliary vector, $\xi_l$  the linear susceptibility of the material, and $\xi_k$  the Kerr susceptibility. The electromagnetic response of the generated plasma is represented by $\textbf{J}_p$ and  laser depletion due to photo ionization (PI) by  $\textbf{J}_{PA}$. 
 The evolution of the free electron density, $n$,  is described by $ \frac{d n}{d t}=W_{PI}(E)$, where $W_{PI}$ is the PI rate. We use Keldysh's formulation for  $W_{PI}$\cite{Keldysh}.  The adiabaticity parameter for solids is $\gamma=\omega_0\sqrt{m_{eff}U_i}/eE$, where $m_{eff}=0.635m_e$ is the reduced mass of the electron and the hole, $U_i$ is the band gap energy, $E$  the laser electric field, and $\omega_0$  the laser angular frequency; for our parameters, $\gamma$ becomes unity for the intensity of $3.5\times10^{13}~ W/cm^2$.
Since we consider intensities higher than this here (or for cases where they are lower, self-focusing causes them to become higher), the multi-photon ionization model is not appropriate. We neglected avalanche ionization as we consider only short ($50~fs$) laser pulses. Our simulation results including avalanche did not differ from simulations without avalanche ionization.\\
 We assume a laser beam focused by a perfectly reflecting parabolic mirror characterized  by  a
given $f\#$, corresponding to laser beam waist sizes of $w_0=1.5,1.0$, and $0.69$~$\mu m$. The laser incident onto the mirror is  a Gaussian beam whose beam waist is half the size of the mirror. To describe the fields focused  by  the  parabolic  mirror,  the  Stratton-Chu integrals\cite{Stratun,Popov2009} are used, which  specify  the  exact  electromagnetic  field emitted  by  the  given  parabolic  surface.  This  field  is  calculated on five boundaries of the 3D FDTD simulation in a total field/scattered field framework. We also considered a laser beam with $w_0=2~\mu m$, where a  paraxial Gaussian beam was used  to save computational resources and time; we have verified that it gives almost exactly the same results as the mirror focused laser beam. The laser pulses are Gaussian in time with a pulse duration of 50 $fs$ and a wavelength of $\lambda = 800~ nm$ and they are linearly polarized along the $y$ direction and propagating along the $x$ direction. The geometrical laser focus is located at $x = 40 ~\mu m$, and the simulation domain is $100 ~ \mu m \times 16 ~\mu m \times 16 ~\mu m$, with
grid size  $\Delta x = \Delta y = \Delta z = 0.02 ~ \mu m$. The background refractive index of silica is $1.45$, the band gap energy is $9 ~eV$, the third order nonlinear susceptibility is $\chi^3 = 1.9x10^-4~ esu$, and the saturation density is  $10n_{cr}$, where $n_{cr}=\frac{m_e \omega_0^2}{4\pi^2}$ is the critical plasma density. \\
\section{Results and discussions}
 \subsection{Effect of laser spot size }
\begin{figure*}[ht]
\includegraphics[width=\linewidth]{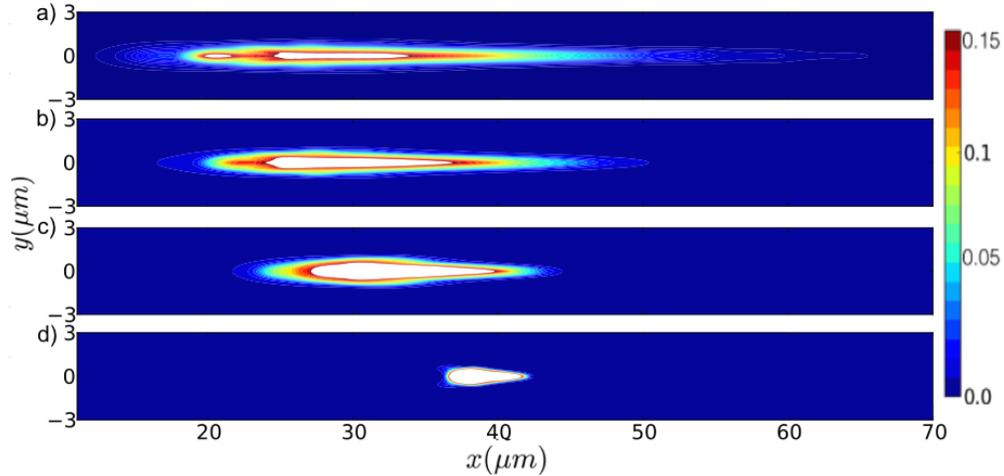}
\caption{Contour plots of electron density after the laser pulse (a-d) for beam waists of $w_0=2.0, 1.5,1.0$, and  $0.69~\mu m$ for a fixed laser energy of $0.32~\mu J$. (see Visualizations 1,2,3 for time-domain movies of laser intensity and free electron density simulations with $w_0=2.0 ,1.0,0.69~\mu m$, respectively.) }\label{fig1}
\end{figure*}
Figure \ref{fig1} shows the contour plots of final electron densities in the $x-y$ plane for simulations performed for different laser focusing conditions corresponding to beam waist sizes of  $w_0=2.0, 1.5, 1.0, 0.69 ~ \mu m$ for a fixed  laser energy of $0.32 \mu J$. Thus, the smaller the spot size of the laser, the higher the  laser intensity would be if the laser propagation were in vacuum. The magnitude of the change of refractive index $\Delta n$ corresponding to permanent damage in fused silica has been measured in the experiments in Ref. \cite{Coua2005}. They found that  permanent damage happens in fused silica when $n >0.15n_{cr}$. Thus, as a rough indication  indication of the permanent damage zones predicted by our simulations, we indicate in white in Fig. \ref{fig1}  (and subsequent figures) the regions in the electron density contour plots corresponding to $n>0.15n_{cr}$.\\
 The transition from  long filamentation for  $w_0>1.0 \mu m$ to a compact structure for tighter focusing conditions is clearly demonstrated in Fig. \ref{fig1}.  For $w_0=2.0~\mu m$ (Fig. \ref{fig1}a), one long filament with multiple damage zones forms, similar in shape and size to the experimental structure reported by Sudrie \textit{et. al.} \cite{Sudrie2002,Coua2005}, proposed at the time that this could be formed from pulse focusing/defocusing/refocusing, but were not able to capture it with their simulations. In Visualization 1, we present a time-domain movie for this simulation that allows us to visualize the dynamics of the laser intensity and the free electron density. We see that  plasma formation begins at about $20~\mu m$, which is $20~\mu m$ before the geometrical focus at $40~\mu m$. As the simulation progresses, the leading half of the pulse experiences a small drop in intensity due to defocusing that leads to a decrease in plasma generation. The tail half of the pulse, however,  visibility deforms around this newly formed plasma due to plasma defocusing. After this, the leading half of the pulse has enough intensity that it refocusses, ultimately creating a longer filament that persists up to and beyond the geometric focus. The tail half of the pulse eventually refocusses as well, though at much lower intensity, and contributes to extending the plasma tail  to $60 ~\mu m$, however, at much lower electron density levels than would be needed for permanent damage.\\
 Figure \ref{fig1} shows that as we decrease the spot size of the laser, the filament
length shortens, ultimately becoming a compact structure with much higher electron density for very tight focusing ($w_0 = 0.69~ \mu m$). One also observes in Fig. \ref{fig1} that for looser focusing, plasma creation begins sooner, much before the geometrical focus, despite the fact that the incident laser intensity is smaller for larger spot sizes (since we keep the total energy constant). This indicates that self focusing and the nonlinear Kerr effect must play an important role in the longer filamentation regime. In contrast, for tighter focusing, the position of the focus is very close to the geometrical focus, indicating a potentially smaller contribution of the nonlinear Kerr effect. 

Visualizations 2,3 show the time domain movies of filament formation for $w_0 =1.0$ and $0.69~\mu m$, respectively. In Visualization 2, we see that plasma formation does not begin until around $27~\mu m$, and that tail half of the pulse again experiences visible plasma defocusing, while the leading edge continues to create plasma until the geometric focus, after which it rapidly defocusses. The tail half does not refocus and thus does not extend the plasma tail, as it did in Visualization 1. In Visualization 3, the tightest focusing we consider, geometric focusing is clearly dominant. The plasma is confined to near the geometric focus, and is rapidly formed. Plasma defocusing is also seen, but since geometric defocusing is so strong (after the geometrical focus), it dominates over Kerr self-focusing, and thus no refocusing is observed.\\
 
 
\begin{figure}[h]
\begin{center}
\includegraphics[width=\linewidth]{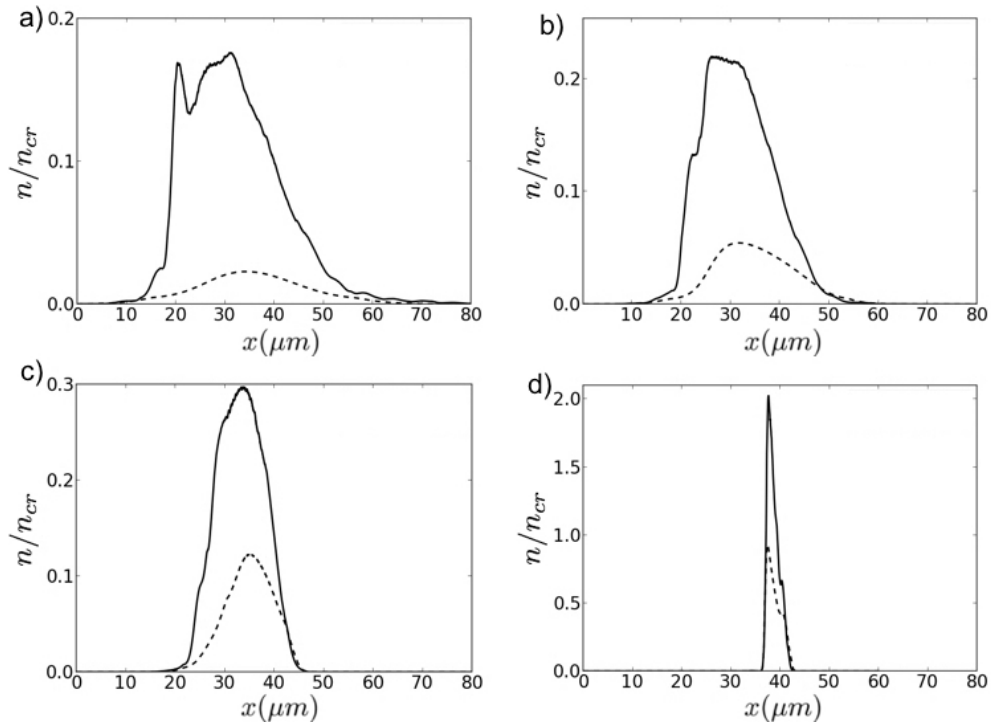}
\end{center}
\caption{On-axis final electron density for $w_0=2~\mu m$ (a),  $w_0=1.5~\mu m$ (b),  $w_0=1~\mu m$ (c), and  $w_0=0.69~\mu m$ (d). The solid lines correspond to the simulations of Fig. \ref{fig1}, whereas the dashed lines correspond to equivalent simulations without the Kerr effect. }\label{fig2}
\end{figure}
 To better understand the role of Kerr self-focusing, we performed simulations equivalent to those of Fig. \ref{fig1}, except that we have turned off the nonlinear Kerr effect by setting the Kerr susceptibility to zero. Figure \ref{fig2} a-d show  the final plasma density along the laser axis  for $w_0=2, 1.5, 1.0, 0.69 ~\mu m$, respectively. The solid lines correspond to the simulations including the nonlinear Kerr effect (as in Fig. \ref{fig1}) and dashed lines correspond to the simulations with Kerr effect turned off. \\
 \begin{figure}
\begin{center}
\includegraphics[width=\linewidth]{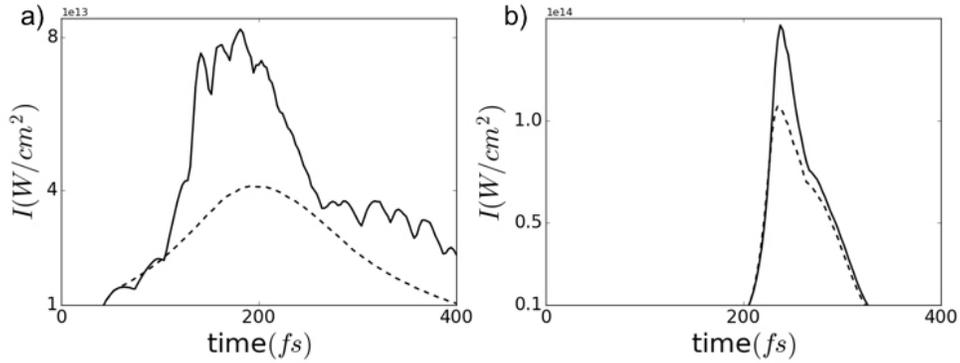}
\end{center}
\caption{Global on-axis maximum intensities as a function of time calculated from simulations  for $w_0=2.0$ (a) and $0.69~\mu m$ (b). Solid and dashed  curves correspond to simulations with and without Kerr nonlinearity, respectively. }
\label{fig3}
\end{figure}
We find that the Kerr effect plays a significant role in all four cases. However, in Fig. \ref{fig2}a and b, corresponding to $w_0 = 2.0$ and $1.5 ~ \mu m$, respectively,  neglecting the Kerr effect causes the plasma density to be far below the threshold for permanent damage. In Fig. \ref{fig2}c, corresponding to $w_0 = 1.0~\mu m$, we find that the electron density does approach a value that could result in permanent damage, but that the damage spot would be much smaller in the absence of Kerr self-focusing. In contrast, in Fig. \ref{fig2}d, for the tightest focusing with $w_0 = 0.69~ \mu m$, we see that while there is still a difference between the two curves, the effective  shape and  length of permanent damage area are very similar. However, the electron density is doubled when the Kerr effect is included, indicating that it plays a key role in the onset of micro-explosions and void formation.\\
   To further investigate the role of the nonlinear Kerr effect, we extracted from our simulations the maximum value that the light intensity reached along the laser axis at each time step. We call this the "global maximum", and we plot this as a function of time in Fig. \ref{fig3} for $w_0= 2.0~\mu m$ (a) and $0.69~\mu m$ (b).  As before, the solid line indicates results corresponding to the simulations of Fig. \ref{fig1}, and the dashed lines the equivalent simulations where the nonlinear Kerr effect is turned off. 
In the solid curve of Fig. \ref{fig3}a, we see an initial  growth of the field intensity due to  self-focusing, which then oscillates around a saturation level after significant plasma is produced. Here we see an interplay between plasma defocusing and nonlinear Kerr self-focusing which ultimately creates the multiple filamentation spots we observe in Fig. \ref{fig1}. In the absence of the nonlinear Kerr effect (dashed curve), this saturation and oscillation behaviour is completely absent.\\
In contrast, we see from Fig. \ref{fig3}b, that for tight focusing, the difference between the simulations with and without the Kerr effect is not as dramatic, and the the light intensity reaches its maximum at the same time for both simulations and follows the same shape. The deviation between the actual values of the maximum  shows that the Kerr effect still does play an important role, however, and is thus likely relevant in determining the nature of the damaged area. The size and shape of this area, however, are primarily determined by  geometrical focusing.\\

\subsection{Analytical Model}
In this section, we demonstrate  more visually the relative importance of Kerr self-focusing versus geometric focusing and how these interact with plasma defocusing by extending to solids, an analytical model previously developed by Lim \textit{et al} \cite{Lim2014}. They used a quantity called wavefront sag $(S)$, which is the path difference between the center and the edge of the wavefront.  For a focusing Gaussian beam, the sag from geometrical focusing is \cite{Lim2014},
\begin{equation}
  S_G=\frac{w_0^2}{2x_R^2}(x-f)
\end{equation}
where $w_0$ is the beam radius, $x$ is longitudinal position, $f$ is the geometrical focus position,  and $x_R=\pi w_0^2/\lambda_0$ is the Rayleigh distance. \\
The  wavelength sags from the Kerr nonlinearity and plasma defocusing are obtained from calculating the  optical path length differences between the wavefronts along the laser axis between the center and the edge of the beam ($1/e^2$ of the intensity at the center),
\begin{equation}
  S=-\int_{0}^{x}\Delta n(x') dx'
\end{equation}
where for the Kerr nonlinearity, we set $\Delta n(x)=n_2 I_0(x)$, where $n_2$ is the nonlinear refractive index,  $I_0(x)=2P_0/\pi w(x)^2$ is the peak laser intensity, $P_0$ is the peak laser power, and $w(x)=w_0\sqrt{1+((x-f)/x_R)^2}$. For plasma defocusing $\Delta n(x)=-\frac{\rho(x)}{2\rho_c}$, where $\rho$ is the on-axis plasma density and $\rho_c$ is the critical plasma density.  Therefore, the sag from Kerr nonlinearity  can be expressed as\cite{Lim2014}:

\begin{equation}
  S_K=\frac{2n_2P_0x_R}{\pi w_0^2}(tan^{-1}\frac{x-f}{x_R}+tan^{-1}\frac{f}{x_R})
\end{equation}
\begin{figure}
\begin{center}
\includegraphics[totalheight=0.18\textheight]{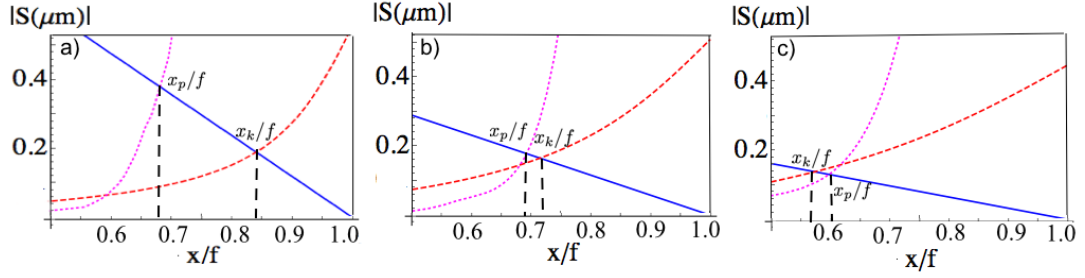}
\end{center}
\caption{ Illustration of calculations of absolute values of sag contribution from Kerr self focusing  ($S_K$, red), plasma defocusing ($S_P$, purple) and geometrical focusing ($S_G$, blue) as a function of normalized distance $(x/f)$ for focusing conditions $w_0=0.69,1.5,2~\mu m$, (a-c) respectively.}\label{sag}
\end{figure}
Lim \textit{et. al}\cite{Lim2014}  used a multi-photon ionization model to calculate the sag from plasma defocusing. However, as we discussed in Sec. \ref{NM} , the multi-photon ionization model is not appropriate here, thus we use instead  the Keldysh ionization rate to obtain,
\begin{equation}
  S_p= -\frac{\tau}{2\rho_c}\int_{0}^{x} W_{PI}(x') dx', 
\end{equation}
where we use numerical integration to calculate $S_p$.\\
 Figure \ref{sag} shows the plot of the contributing wavelength sags for three simulations of Fig. \ref{fig1} ($w_0=0.69, 1.5, 2.0~\mu m$). For all three cases, when the beam is far from the geometrical focus, the intensity is low, thus the nonlinear Kerr effect and plasma defocusing are weak and  $|S_G| $(blue) is much larger than $|S_K|$(red) and $|S_p|$(purple). The position $x_k$ is defined as the position where $|S_g|=|S_K|$, where Kerr self focusing becomes non-negligible compared to geometrical focusing. Similarly, $x_p$ is the position where $|S_g|=|S_p|$, where plasma defocusing becomes non-negligible.
In the tightest focusing regime (Fig. \ref{sag}a),  geometrical focusing and  plasma defocusing are the primary contributions ($x_k>x_p$).  For $w_0=1.5~\mu m$ (Fig. \ref{sag}b), $x_k$ is closer to $x_p$, therefore the Kerr nonlinearity and plasma defocusing are comparable and both play important roles. However, in the loosest focusing regime ($w_0=2~\mu m$, Fig. \ref{sag}c), the Kerr effect builds up faster than the plasma defocusing and geometrical focusing  ($x_p>x_k$), therefore the Kerr nonlinearity plays the primary role. This is in agreement with our simulation results in the previous subsection.\\
 \subsection{The effect of laser energy for fixed laser spot sizes}
 \begin{figure*}[ht]
\includegraphics[width=\linewidth]{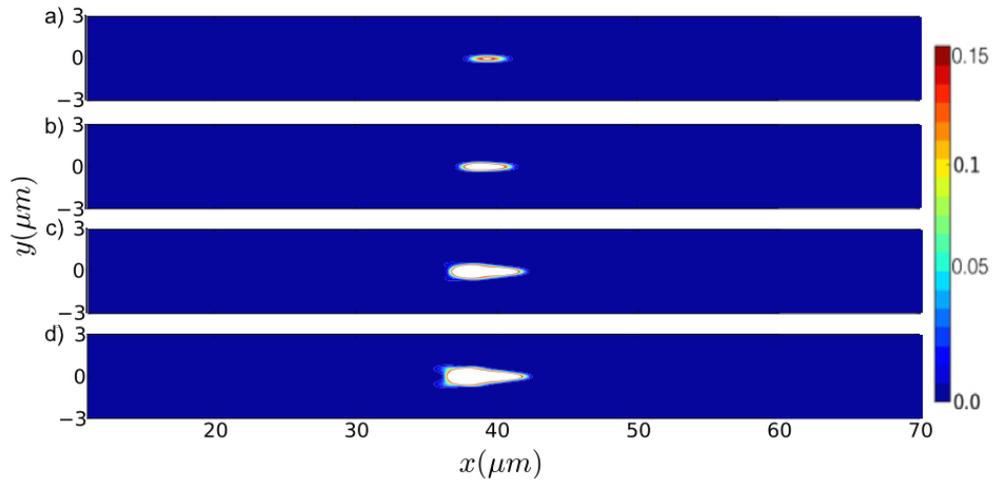}
\caption{Contour plots of electron density for $w_0=0.69~\mu m$ for intensities of $I=5e13$ (a), $1.1e14$ (b), $3.3e14$ (c), and $4.9e14~W/cm^2$ (d).}\label{fig5}
\end{figure*}
\begin{figure}
\begin{center}
\includegraphics[totalheight=0.3\textheight]{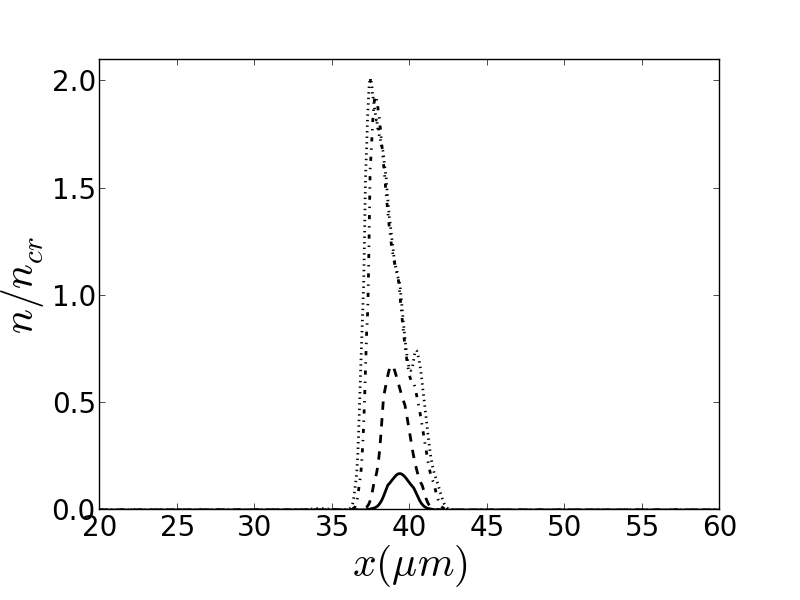}
\end{center}
\caption{On-axis electron density along the laser propagation direction $x(\mu m)$ from simulations  for $w_0=0.69~\mu m$ for $I= 5e13$ (solid line), $1.1e14$ (dashed line), $3.3e14$ (dashed-dotted line), and $4.9e14$ (dotted line) $~W/cm^2$.}
\label{fig5a}
\end{figure}
 \begin{figure*}[ht]
\begin{center}
\includegraphics[width=\linewidth]{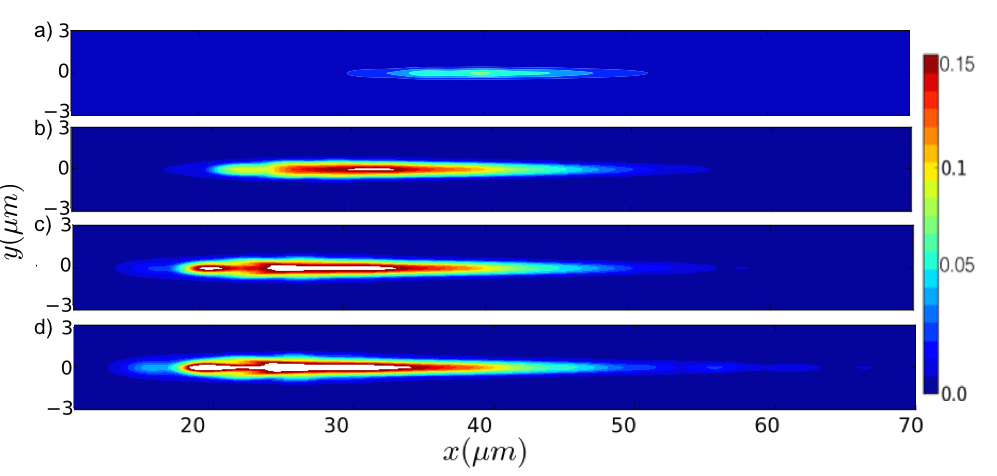}
\end{center}
\caption{Contour plots of electron density for $w_0=2~\mu m$ for $I=2e13$ (a), $4e13$ (b), $5e13$ (c), and  $6e13$ (d) $~W/cm^2$.}\label{fign}
\end{figure*}
\begin{figure}
\begin{center}
\includegraphics[totalheight=0.3\textheight]{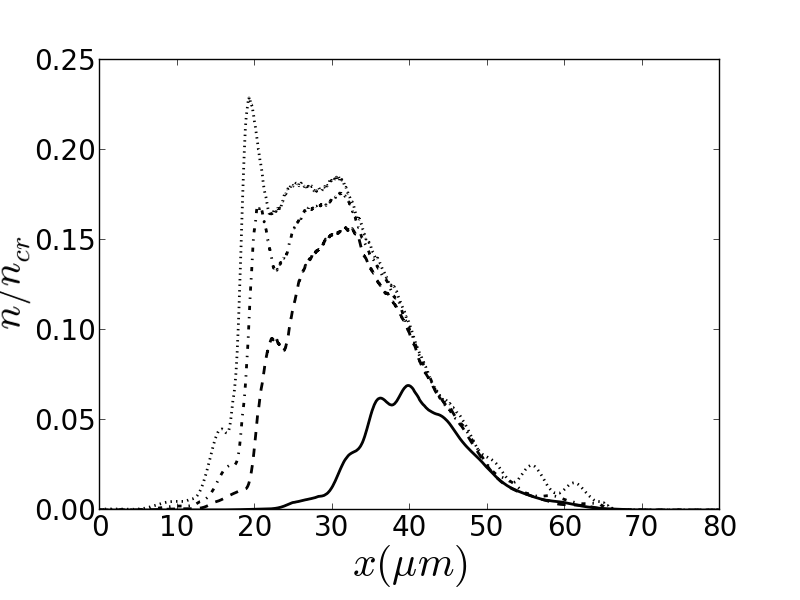}
\end{center}
\caption{On-axis electron density along propagation direction $x(\mu m)$ from simulations  for $w_0=2~\mu m$ for  $I=2e13$ (solid line), $4e13$ (dashed line), $5e13$ (dashed-dotted line), and $6e13$ (dotted) $~W/cm^2$.}
\label{fig6}
\end{figure}
 In what follows we study how the input laser energy for fixed laser spot size (and fixed laser pulse duration) affects the interaction with  bulk fused silica. Figure \ref{fig5} shows  electron density contour plots after the laser pulse for $w_0=0.69~\mu m$, for peak incident laser intensities of: $5\times 10^{13}, 1.1 \times 10^{14}, 3.3 \times 10^{14}, 4.9\times 10^{14}~ W/cm^2$ (Fig. \ref{fig5}a-d, respectively).  We observe that when the laser peak intensity is varied,  the focus position does not change considerably. 
 
 We found that the threshold for permanent damage happens for a laser peak intensity of  $5 \times 10^{13}~W/cm^2$  which leads to a plasma size of  $\approx 1~\mu m$, with peak plasma density $0.16n_{cr}$. Increasing the laser intensity to $1.1 \times 10^{14}~W/cm^2$ leads to a longer ($\approx 2.9~\mu m$)  oval shape structure with maximum electron density $0.65n_{cr}$.  The damage area for laser peak intensity of $3.3\times 10^{14}~W/cm^2$ is elongated ($\approx 4.6~\mu m$) and has a pear shape structure the same as increasing the peak laser intensity to $4.9\times 10^{14}~W/cm^2$ leads to very similar structure as  $3.3\times 10^{14}~W/cm^2$. \\
Figure \ref{fig5a} shows the on-axis values of the electron densities corresponding to the simulations of Fig. \ref{fig5}. While the plasma shape elongates as the intensity is increased, and the plasma density increases, we see a saturation in the electron density for $3.3 \times 10^{14}~W/cm^2$ and above.\\
 Figure \ref{fign}a-d  shows the electron density distribution for a larger laser spot size ($w_0=2~\mu m$) for varying  laser pulse energy, with peak intensities of  $I=2\times 10^{13}, 4\times 10^{13},5\times 10^{13}$, and $ 6\times 10^{14}~W/cm^2$, respectively. The corresponding electron densities along the laser axis are plotted in Fig. \ref{fig6}.\\ 
 An area of very low electron density with maximum density $0.06n_{cr}$ is formed for $I=2 \times 10^{13}~W/cm^2$, which is well below damage threshold density.  $I=4\times 10^{13}~W/cm^2$ is the threshold for filament formation, with a filament length of around $4.2~\mu m$. As we increase the laser energy (and thus the laser intensity), the location where plasma is first created moves backward, as expected, and  the length of total damaged  area increases. Further, as  intensity  increases, we see the emergence of a prominent peak on the left, which gets larger with increasing intensity relative to those on the right. This trend suggests that as laser energy increases, we would expect two different type of damage zones, with the left zone being smaller and containing more structural damage, and the right zone much more elongated, perhaps reaching the threshold for permanent refractive index change but not structural damage. This agrees well with the  observations described in Refs.\cite{Sudrie2002,Coua2005}  
 \section{Conclusion}
 In this paper we studied the mechanisms underlying intense laser interaction and filamentation in fused silica at 800 nm by performing 3D high resolution FDTD simulations, thus bringing new insight into experiments going back two decades. We carried out simulations for a range of focusing conditions, including tight focusing, and a range of laser energy. As expected, we found that filamentation strongly depends on laser focusing, and that Kerr self-focusing plays a key role in each regime. For looser focusing, it is the primary mechanism determining the shape and strength of the created plasma, and we observed a refocusing of the laser pulse after initial plasma creation and defocusing, leading to additional plasma creation and multiple damage zones. For tight laser focusing, while the compact shape of the created plasma is determined by geometrical focusing, Kerr self focusing has a large effect on the maximum electron density attained within the plasma, signifying its important role in  the onset of structural damage and void creation through mirco-explosions. Both of these scenarios become more pronounced with an increase in laser energy.    

\section{Acknowledgement}
This work was supported  by the Canada Research Chairs program, the Canadian Foundation for Innovation, and  the U.S. AFOSR FA9550-14-1-0247. The authors also wish to thank West-grid and the Compute Canada consortium for computational resources.
\section{Disclosures}
The authors declare no conflicts of interest.

\end{document}